\documentclass[a4paper,twocolumn]{esapub2005} 
\def\zerozero  {SGR~1900$+$14}
\def\zerosei  {SGR~1806--20}
\def\int {{\it INTEGRAL}}

\usepackage{times}
\usepackage{natbib}
\usepackage{graphicx}

\title{INTEGRAL and Magnetars}
\author[1]{Diego G\"otz}
\affil[1]{CEA, DSM/DAPNIA/Service d'Astrophysique, Gif-sur-Yvette, France}
\author[2]{Sandro Mereghetti}
\affil[2]{INAF -- Istituto di Astrofisica Spaziale e Fisica Cosmica, Milano, Italy}
\author[3]{Kevin Hurley}
\affil[3]{University of California at Berkeley, Space Sciences Laboratory, Berkeley CA, USA}
\author[4]{I. F\'elix Mirabel}
\affil[4]{European Southern Observatory, Santiago, Chile}
\author[2]{Paolo Esposito}
\author[2]{Andrea Tiengo}
\author[5]{Georg Weidenspointner}
\affil[5]{Centre d'\'Etude Spatiale des Rayonnements, Toulouse, France}
\author[6]{Andreas von Kienlin}
\affil[6]{Max-Planck-Institut f\"ur extraterrestrische Physik, Garching, Germany}

\begin{document}


\maketitle

\begin{abstract}
Thanks to \int's long exposures of the Galactic Plane, the two
brightest Soft Gamma-Ray Repeaters, SGR 1806-20 and SGR 1900+14, have been
monitored and studied in detail for the first time at hard-X/soft-gamma rays. 

SGR 1806-20, lying close to the Galactic Centre, and being very active
in the past two years, has provided a wealth of new \int~ results,
which  we will summarise here: more than 300 short bursts have been observed
from this source and their characteristics have been studied with
unprecedented sensitivity in the 15-200 keV range. A hardness-intensity
anticorrelation within the bursts has been discovered and the overall Number-Intensity
distribution of the bursts has been determined. The increase of its
bursting activity eventually led to the December 2004 Giant Flare for
which a possible soft gamma-ray ($>$80 keV) early afterglow has been detected with \int.

The deep observations allowed us to discover the persistent emission in
hard X-rays (20-150 keV) from 1806-20 and 1900+14, the latter being in
quiescent state, and to directly compare the spectral characteristics
of all Magnetars (two SGRs and three Anomalous X-ray Pulsars) detected
with \int. 
\end{abstract}

\section{Introduction}
Most neutron stars (NSs) belong to two big categories: isolated NSs, where the dominant source of energy is the rotational one, and binary NSs where the accretion process dominates. The first category is mainly represented by classical radio pulsars with spin periods between 1.5 ms and 5 s, and comprises more than 1500 members. The youngest among them are also detected at higher energy. The second category is composed by several hundreds of members, which are hosted in binary systems and classified, based on the mass of their companion, as low mass (LMXB) or high mass X-ray binaries (HMXB).
Many among them are transients and their rotational periods range from a few ms (the so called {\it recycled} millisecond pulsars) to a few hours (the longest one, $\sim$5 hours, recently discovered in 4U 1954+319, \cite{mattana}).

Besides rotation and accretion two other sources of energy can dominate: in the ``middle aged" isolated NSs thermal emission from the surface is detected in X-rays, and in LMXBs nuclear energy can be the dominant source, since they sometimes emit the so called type I X-ray bursts, whose flux reaches the Eddington limit for a NS.

Magnetars do not fit in any of the above categories. They are a peculiar class of sources, made of 11 members and 4 candidates, where the magnetic energy is the dominant one. This class is composed by Soft Gamma-Ray Repeaters (SGRs) and Anomalous X-ray Pulsars (AXPs): the reason for the different names is historical and due to the fact that they were discovered through different manifestations.

The SGRs were discovered more than 20 years ago, thanks to the fact that they emit short ($\sim$ 0.1) and intense (10$^{39}$-10$^{42}$ erg s$^{-1}$) bursts of  high-energy radiation in the tens to $\sim$hundred keV energy range. They were initially considered a subclass of the Gamma Ray Bursts (GRBs), but then their recurrent behaviour and their soft spectrum (compared to GRBs) identified them as a separate class of objects.
The rate of burst emission in SGRs is highly variable. Bursts are generally emitted during
sporadic periods of activity, lasting days to months, followed by
long ``quiescent'' time intervals (up to years or decades) during
which no bursts are emitted. Occasionally SGRs emit ``giant
flares'', that last up to a few hundred seconds
and have peak luminosity up to  10$^{46}$-10$^{47}$ erg s$^{-1}$. 
Only three giant flares have been observed to date, each one from
a different source, see Tab. \ref{tab:gf} (see, e.g., \cite{mazets} for 0526--66, \cite{hurley1999} for 1900+14,
\cite{swiftgiant,acsgiant,rhessigiant} for 1806--20). 
\begin{table}[ht!]
\caption{Energetics of the three giant flares detected to date.}
\vspace{1em}
\begin{tabular}{|c|c|c|}
\hline SGR & Initial Pulse Energy [ergs] & Tail energy [ergs] \\ 
\hline 0526-66 & 1.6$\times$10$^{44}$ & 4$\times$10$^{44}$ \\ 
\hline 1900+14 & $>$7$\times$10$^{43}$ & 5$\times$10$^{43}$\\ 
\hline 1806-20 & 2$\times$10$^{46}$ &  10$^{44}$\\ 
\hline 
\end{tabular} 
\label{tab:gf}
\end{table}
These giant flares are today the most convincing
evidence for the existence of very high ($B\sim$10$^{15}$ G) magnetic fields associated with these objects \cite{dt92,pac92,td95}, dubbed {\em Magnetars}. The pulsations measured during these powerful flares
indicated for the first time the NS nature of these sources. The same periodicities (5--8 s) have also been confirmed 
in the persistent (i.e. non-bursting) emission, which is observed from SGRs
in the soft X--ray range ($<$10 keV) \cite{kouve98,kh1900,woods99}, with a typical luminosity of $\sim$10$^{35}$ erg s$^{-1}$. Spindown rates,
$\dot P$, of the order of 10$^{-11}$ s s$^{-1}$ have been measured in these objects.

AXPs were identified by \citet{merestella} as a subclass of LMXBs. Later studies (see \citep{mere02} for a review) demonstrated that AXPs show no evidence for companion stars, having faint IR counterparts and lacking of orbital Doppler delays in their light curves. Two (maybe three) of them are associated with Supernova remnants, and recent evidence has been put forward for the existence of a fallback disk in 4U 0142+61 \cite{wang}.
 Their rotational periods are clustered in the 5--12 s range with a secular spindown of 0.05-4$\times$10$^{-11}$s s$^{-1}$. Their X-ray spectra (1-10 keV) are soft and usually modelled with the sum of a black body  ($kT\sim$ 0.5 keV) and a power law (photon index, 2$<\Gamma<$4), and the typical luminosities are in the range 10$^{34-36}$ erg s$^{-1}$. This luminosity is much larger than the spindown one 
($I\omega\dot\omega$), indicating that the dominant source of energy in AXPs, as in SGRs, is the magnetic one. In fact given the period and period derivative values, one expects a surface dipole magnetic field of

\begin{equation}
B=\left(\frac{3Ic^{3}P\dot P}{2\pi^{2}R^{6}}\right)\simeq3.2\times10^{19}(P\dot P)^{1/2} \; \rm G, 
\end{equation}

where $I$ ($\simeq$10$^{45}$g cm$^{2}$) is the NS moment of inertia, and $R$ ($\simeq$10$^{6}$ cm) is the NS radius. The derived field values for AXPs exceed the quantum critical value of $B_{Q}\equiv m_{e}^{2}c^{3}/(e\hbar)=4.4\times10^{13}$ G, and hence AXPs have been proposed to be also members of the Magnetar class even if this model had been developed earlier for explaining the SGR phenomenology.
 
The unification process of the two categories started on the base of the similar temporal and spectral behaviour at soft X-rays. But the real breakthrough was the detection of SGR-like bursts from four AXPs (1E 1048.1-5937, \cite{gavrill}, 1E 2259+586, \cite{kaspi}, XTE J1810--197, \cite{woods}, 4U 0142+61, \cite{kaspi4u}).
The Magnetar model (see \cite{dt92,pac92,td95,tlk} for details), predicts that, if the ``proto-NS" is initially spinning at $\sim$few ms, an efficient $\alpha-\Omega$ dynamo process can produce magnetic fields of the order of B$\sim$10$^{15}$ G. Then the huge magnetic field can easily spin down the neutron star  via dipole magnetic braking to periods longer than 10 s in $\sim$10$^{4}$  years. Such a magnetic field also provides the necessary energy to power the spectacular giant flares in SGRs (at least once in an SGR lifetime), magnetically confine the trapped fireball seen in flares' tails, and power the persistent X-ray luminosity.

The new \int~ results concerning SGRs are presented in the following sections. For a review on the low-energy (0.1--10 keV) XMM observations of SGRs see \cite{mere2006}.

\section{\zerosei}
\label{sec:zerosei}
\zerosei~ was  discovered by the Interplanetary Network (IPN) in 1979 \cite{1806disc}. It lies in a crowded region
close to the galactic centre. \citet{kouve98} discovered a quiescent X-ray pulsating ($P$=7.48 s) counterpart, which
was spinning down rapidly ($\dot P$=2.8$\times$10$^{-11}$ s s$^{-1}$). If this spindown is interpreted
as braking by a magnetic dipole field, its strength is $B\sim$10$^{15}$ G. The source activity is variable, 
alternating between quiet periods and very active ones.

After a period of quiescence, \zerosei~ became active in the Summer of 2003  \cite{hurley2003}.
Its activity then increased in 2004, and
a strong outburst during which about one hundred short bursts were emitted in a few minutes
occurred on October 5 2004 \cite{gotz1806}.
Finally a giant flare,
whose energy (a few 10$^{46}$ erg) was two orders of magnitude larger than those of the
previously recorded flares from SGR 0526-66 and SGR 1900+14,
was emitted on December 27$^{th}$ 2004 (see e.g. \cite{swiftgiant,acsgiant,rhessigiant}).

\begin{figure}[ht!]
\centering
\includegraphics[width=8cm]{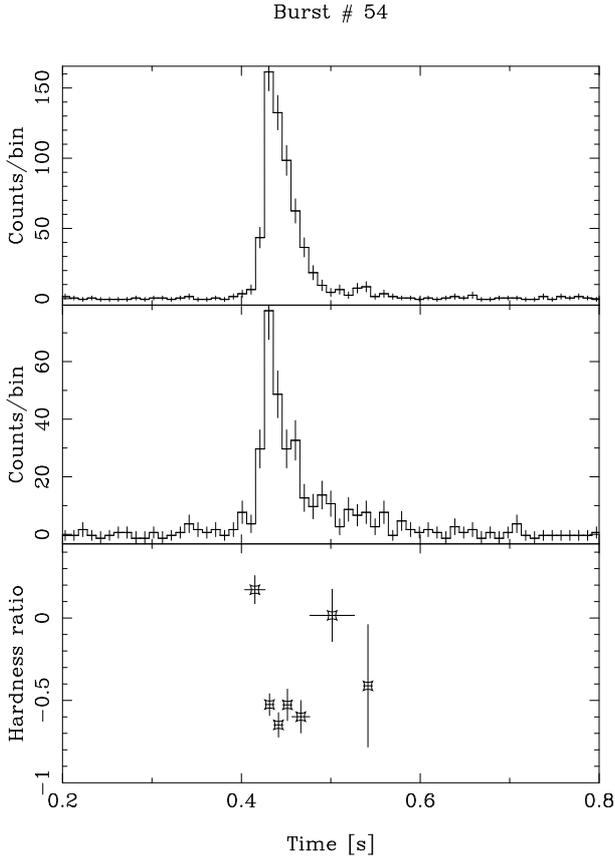}
\caption{IBIS/ISGRI light curves in the soft (20-40 keV, upper
panel) and hard (40-100 keV, middle panel) energy range and
hardness ratio (lower panel) for a short burst from SGR 1806--20.}
\label{fig:evol}      
\end{figure}

\begin{figure}[ht!]
\centering
\includegraphics[width=8.5cm]{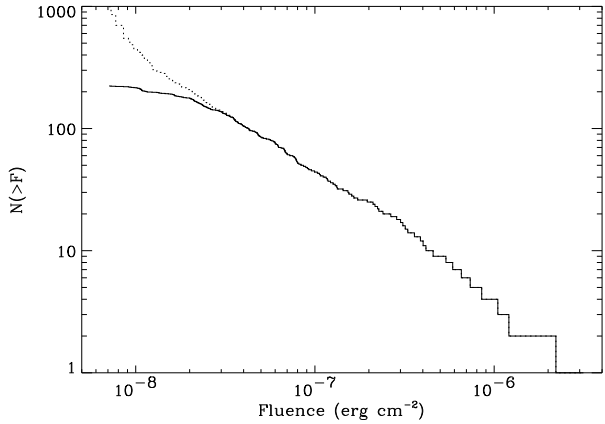}
\caption{Number-intensity distribution of all the bursts detected by \int~ in 2003 and 2004.
The continuous line represents
the experimental data, while the dashed line represents the data corrected
for the exposure. From \cite{gotz1806}.}
\label{fig:logn}
\end{figure}

Thanks to its good sensitivity \int~ detected more than 300 short bursts from \zerosei. Many of these bursts were among the faintest ever detected from these sources, and only thanks to imaging they could be studied in detail for the first time. In fact, \int~ provided new results on them, like the discovery of a hardness-intensity anti-correlation within the bursts, see Fig. \ref{fig:evol}, and the determination of the overall Number-Intensity distribution, see Fig. \ref{fig:logn}, which can be modeled as a single power law ($\alpha$=0.91$\pm$0.09) over 2.5 energy decades. These results have been presented in detail elsewhere 
\cite{gotz04,gotz1806}.
We will focus here on the \zerosei~ persistent emission and on the Giant Flare of December 2004.

\subsection{Discovery of the persistent emission}
Until 2004 spectral information on the perisistent emission
of SGRs was known only below 10 keV, where the spectrum is usually well described by 
the sum of a power law component and a black body (see e.g. \cite{xmm1806}).
In 2005 two groups reported independently the discovery with \int~ of persistent hard X-ray emission originating from \zerosei~ \cite{mere05,molkov}.

\begin{figure}[ht!]
\centering
\includegraphics[width=6.cm,angle=90]{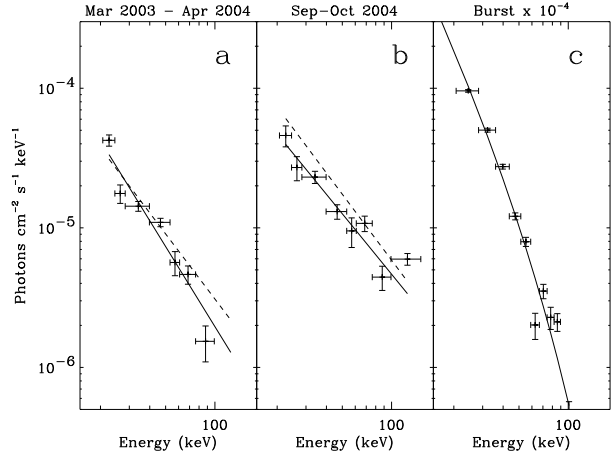}
\caption{IBIS/ISGRI spectra of \zerosei. a) persistent emission March 2003-April 2004, b) persistent emission September-October 2004, c) one burst (scaled down by a factor 10$^{4}$). The solid lines are the best fits (power laws in a) and b), thermal bremsstrahlung in c)). The dashed lines indicate the extrapolation of power-law spectra measured in the 1--10 keV band with {\em XMM-Newton} \cite{xmm1806}. From \cite{mere05}.}
\label{fig:spectra}
\end{figure}

\begin{figure}[ht!]
\centering
\includegraphics[width=6.cm,angle=-90]{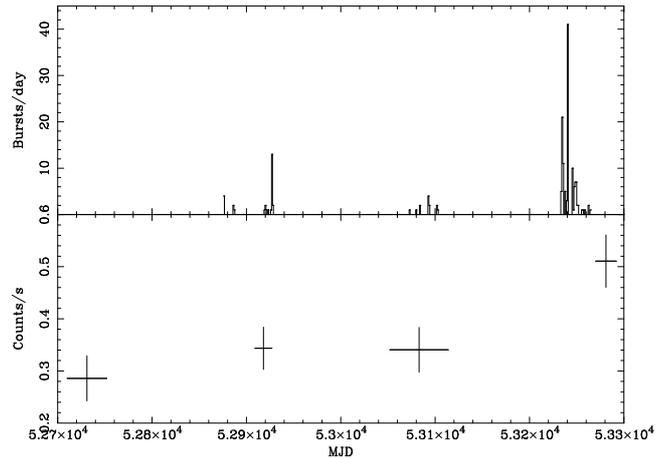}
\caption{Upper Panel: histogram of the number of bursts per day detected with the third Interplanetary Network (IPN). Lower Panel: IBIS/ISGRI count rate in the 20-60 keV band. From \cite{mere05}.}
\label{fig:persburst}
\end{figure}

\begin{figure}[ht!]
\centering
\includegraphics[width=6cm,angle=-90]{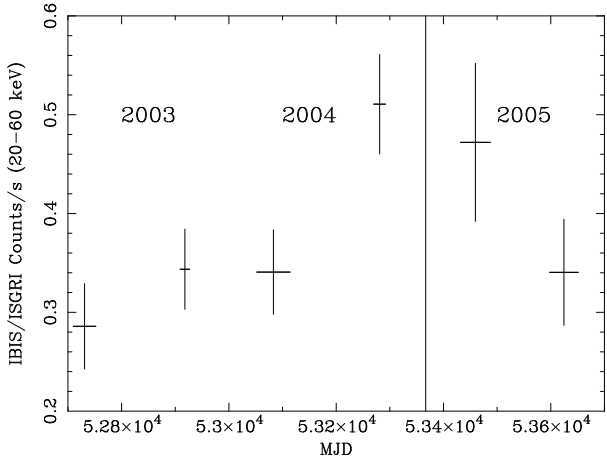}
\caption{Long term light curve of \zerosei, as measured with IBIS. The vertical 
line represents the time of the giant flare of December 27 2004.}
\label{fig:zerosei}
\end{figure}

\int~ showed that the spectrum above 20 keV is rather hard, with a photon index between 1.5 and 2.0 (see Fig. \ref{fig:spectra}) and extends up to 150 keV without an apparent cutoff. 
It connects rather well with the low energy ($<$ 10 keV) spectrum \cite{xmm1806}, 
and the intensity and spectral hardness are correlated with the degree of bursting activity of 
the source \cite{mere05,gotz1806} (see Fig. \ref{fig:persburst}) and with the infrared flux \cite{giallo}. 
Our group is continuously monitoring the hard X-ray flux of \zerosei, and the 
long term light curve of the source is shown in Fig. \ref{fig:zerosei}. As can be seen, the persistent 
flux increased in 2003 and 2004 up to the giant flare (which is marked with a vertical line in the plot), and 
then decreased in 2005.

These results support the twisted magnetosphere model developed by \citet{tlk}, in which a twisted internal magnetic field provides source for helicity of the magnetosphere by shearing the NS crust. The currents, present in the twisted magnetosphere, produce hard spectral tails by resonant cyclotron scattering. In turn the twisted field produces stronger braking than a simple dipole, increasing the spindown rate, as measured by {\em XMM} in \zerosei~ \cite{xmm1806}. At the same time stresses in the crust increase, causing a higher rate of bursts. In the end the giant flare is associated with a major magnetic field reconfiguration, which produces changes in the light curve and spectrum of the SGRs (see below).


\subsection{The Giant Flare of December 27 2004}
A giant flare from \zerosei~ was
discovered with the \int~ gamma-ray observatory on 2004
December 27 \cite{flare}, and detected with many other satellites (e.g. \cite{swiftgiant,rhessigiant,tera}).
The analysis of the SPI-ACS data ($>$80 keV) of the flare, presented in \cite{acsgiant}, showed that
the giant flare is composed by 3 components: an initial spike lasting 0.2 s, followed by a $\sim$400 s 
long pulsating tail, modulated at the neutron star period of 7.56 s, and a possible high-energy afterglow. 
The initial spike was so bright that it saturated the ACS, so we could derive only a lower limit on its fluence,
which turned out to be two orders of magnitude brighter (10$^{46}$ ergs, see e.g. \cite{tera}) 
than the previously observed giant flares from
\zerozero~ \cite{hurley1999}, and SGR 0526--66 \cite{mazets}. The energy contained in the tail (1.6$\times$10$^{44}$ 
ergs), on the other hand, was of the same order as the one in the pulsating tails of the previously observed giant 
flares, see Table \ref{tab:gf}.

\begin{figure}[ht!]
\centering
\includegraphics[width=8.cm]{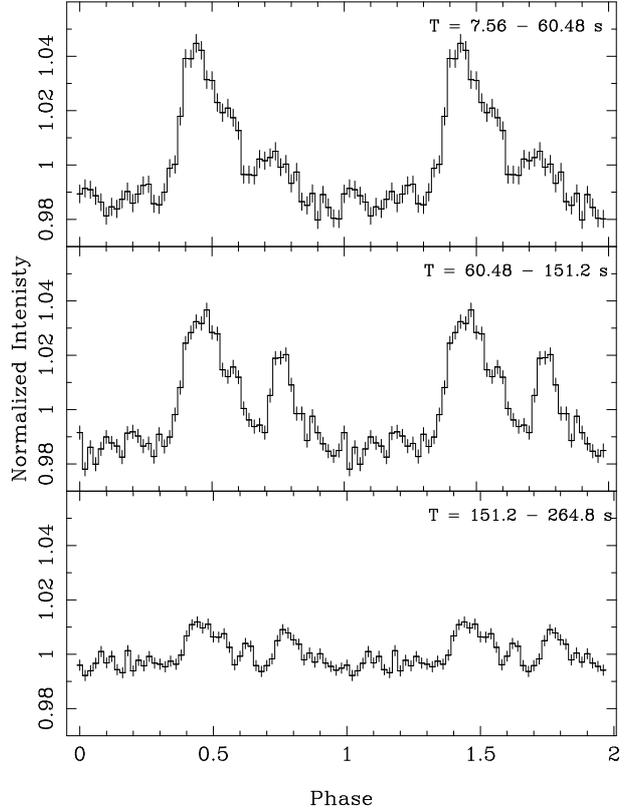}
\caption{Averaged pulse profile of the tail of the giant flare, obtained by folding the data in three time intervals at spin period of 7.56 s. From \cite{acsgiant}}
\label{fig:pulse}
\end{figure}

By folding the ACS light curve at the best period for several cycles, one can appreciate how the pulse profile changes with time during the giant flare itself, see Fig. \ref{fig:pulse}, indicative of changes in the magnetic field configuration.

\begin{figure}[ht!]
\centering
\includegraphics[width=6.1cm,angle=-90]{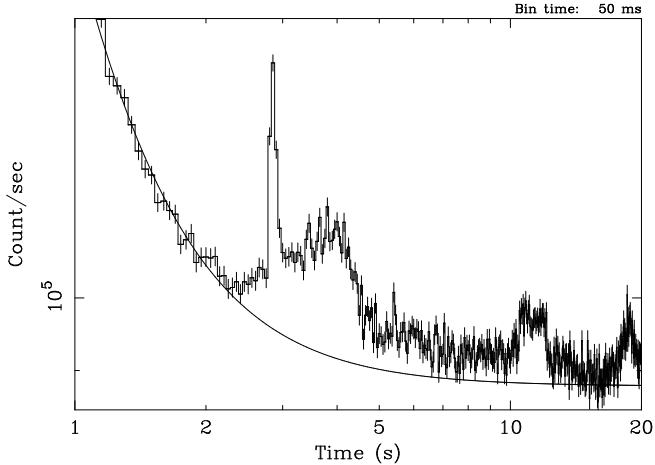}
\caption{SPI-ACS light curve of the giant flare of \zerosei, binned at 50 ms. The Moon reflection component is clearly visible at t$\sim$ 2.8s. Note that the initial part of the flare is not shown.}
\label{fig:moon}
\end{figure}

A $\sim$0.2 s long small burst was detected in the ACS data 2.8 s after the initial spike, see Fig. \ref{fig:moon}.
It is superposed on the 
pulsating tail and has no clear association with the pulse phase. This burst is produced by 
the reflection by the Moon of the initial spike of the giant flare. In fact this delay corresponds to the light 
travel time between \int, the Moon, and back. A similar detection was reported with the {\it Helicon-Coronas-F} 
satellite \cite{mazets05}.

\begin{figure*}[ht!]
\centering
\includegraphics[width=13cm]{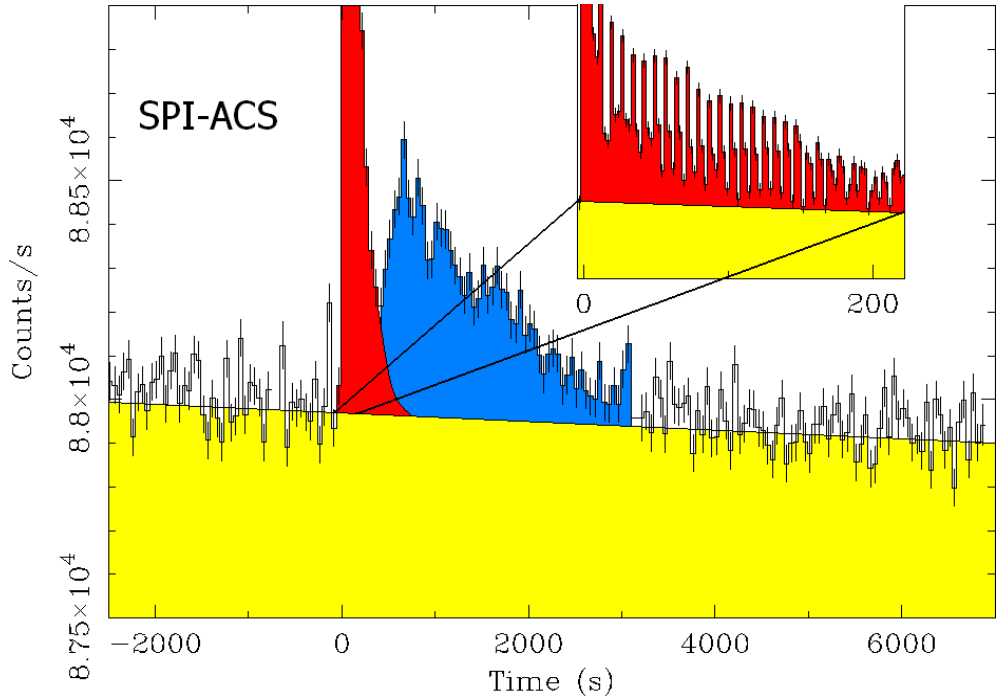}
\caption{Light curve of the Giant Flare of December 27 2004 as measured with 
SPI-ACS above 80 keV. The light curve is binned at 50 s, and hence the pulsating tail
is not visible (it is visible in the inset where the light curve is binned at 2.5 s).
(yellow: instrumental background, red: Flare tail, blue: high-energy afterglow, see text)}
\label{fig:flare}
\end{figure*}

The most striking feature provided by the \int~ data is the detection of a possible early high-energy afterglow 
emission associated with the giant flare. At the end of the pulsating tail the count rate increased again, forming 
a long bump which peaked around t$\sim$700 s and returned to the pre-flare background level at t$\sim$3000-4000. This
component decays as $\sim t^{-0.85}$, and is shown in blue in Fig. \ref{fig:flare}, while the overall long term background trend is shown in yellow,
and the giant flare itself in red. The association of this emission with \zerosei~ is discussed
in \cite{acsgiant}. The fluence contained in the 400-4000 s time interval is approximately the same as that in the
pulsating tail. With simple gamma-ray burst afterglow models based on synchrotron emission one can derive
the bulk Lorentz factor $\Gamma$ from the time $t_{0}$ of the afterglow 
onset: $\Gamma\sim$15($E$/5$\times$ 10$^{43}$ ergs)$^{1/8}$($n$/0.1 cm$^{-3}$)$^{-1/8}$($t_{0}$/100)$^{-3/8}$,  
where $n$ is the ambient density. This is consistent with the mildly relativistic outflow inferred from the radio 
data \cite{granot}.

\section{\zerozero}
\zerozero~ was discovered in 1979 \cite{1900discovery} when it emitted 3 bursts in 2 days.
Since then short bursts were observed from this source with BATSE, {\it RXTE} and
Interplanetary Network satellites in the years 1979-2002. \zerozero~ emitted a
giant flare on August 27 1998 (e.g. \cite{hurley1999}), followed by less intense
``intermediate'' flares on August 29 1998 \cite{ibrahim} and in April 2001 \cite{lenters}. The last bursts
reported from \zerozero~ were observed with the Third Interplanetay
Network (IPN) in November 2002 \cite{ipn1900}. No bursts from
this source were revealed in all the \int~ observations from
2003 to 2005, but Swift has detected renewed activity in 2006 \cite{swift1900}.

\subsection{Discovery of the persistent emission}

Using 2.5 Ms of \int~ data, \citet{gotz1900} reported the discovery of persistent hard X-ray emission. 
This emission extended up to $\sim$ 100 keV, but with a softer spectrum 
compared to \zerosei, having a photon index of 3.1$\pm$0.5. Also the luminosity is dimmer in this case, being
$\sim$4$\times$10$^{35}$ erg s$^{-1}$, a factor of three lower than \zerosei. This is probably due to the fact that \zerozero~ was observed in a quiescent state. In fact a possible detection of a high-energy component in {\em SAX} PDS data is reported by \citet{esposito06} during an active state of the source after the giant flare of August 1998. In that case the hard tail was brighter and harder ($\Gamma\sim$1).
The \int~ observations spanned 
March 2003 to June 2004, and did not include the recent reactivation of the source in March 2006 
\cite{swift1900}, when the source emitted a few tens of regular bursts plus an intense burst series, 
lasting $\sim$30 s \cite{1900outburst}, reminiscent of the October 5 2004 event from \zerosei.
We recently analysed the \int~ data spanning from August 2004 to March 2006, and found that the
hard X-ray flux of the source flux did not increase up to a few weeks before its reactivation.
This indicates that the reactivation was not triggered by a flux increase, at least on the time
scale of a few months sampled by \int.

The soft and constant spectrum of \zerozero ~is possibly related to the fact that
this source is still in a rather quiescent state.

\section{Comparison with the Anomalous X-ray Pulsars}

Hard X-ray persistent emission ($>$20 keV) has recently been
detected with \int~ also from the Anomalous X-ray Pulsars.
It has been detected from three AXPs with
\int: 1E 1841--045 \cite{molkovaxp}, 4U 0142+61
\cite{denhartog} and 1RXS J170849--400910 \cite{revnitsev}.
The presence of pulsations seen with RXTE up to $\sim$200 keV in 1E
1841--045 \cite{kuiper} proves that the hard X-ray emission
originates from the AXP and not from the associated supernova
remnant Kes 73. The discovery of (pulsed) persistent hard X-ray
tails in these three sources was quite unexpected, since below 10
keV the AXP have soft spectra, consisting of a blackbody-like
component (kT$\sim$0.5 keV) and a steep power law (photon index
$\sim$3--4).

\begin{figure}[ht!]
\centering
\includegraphics[width=8.8cm]{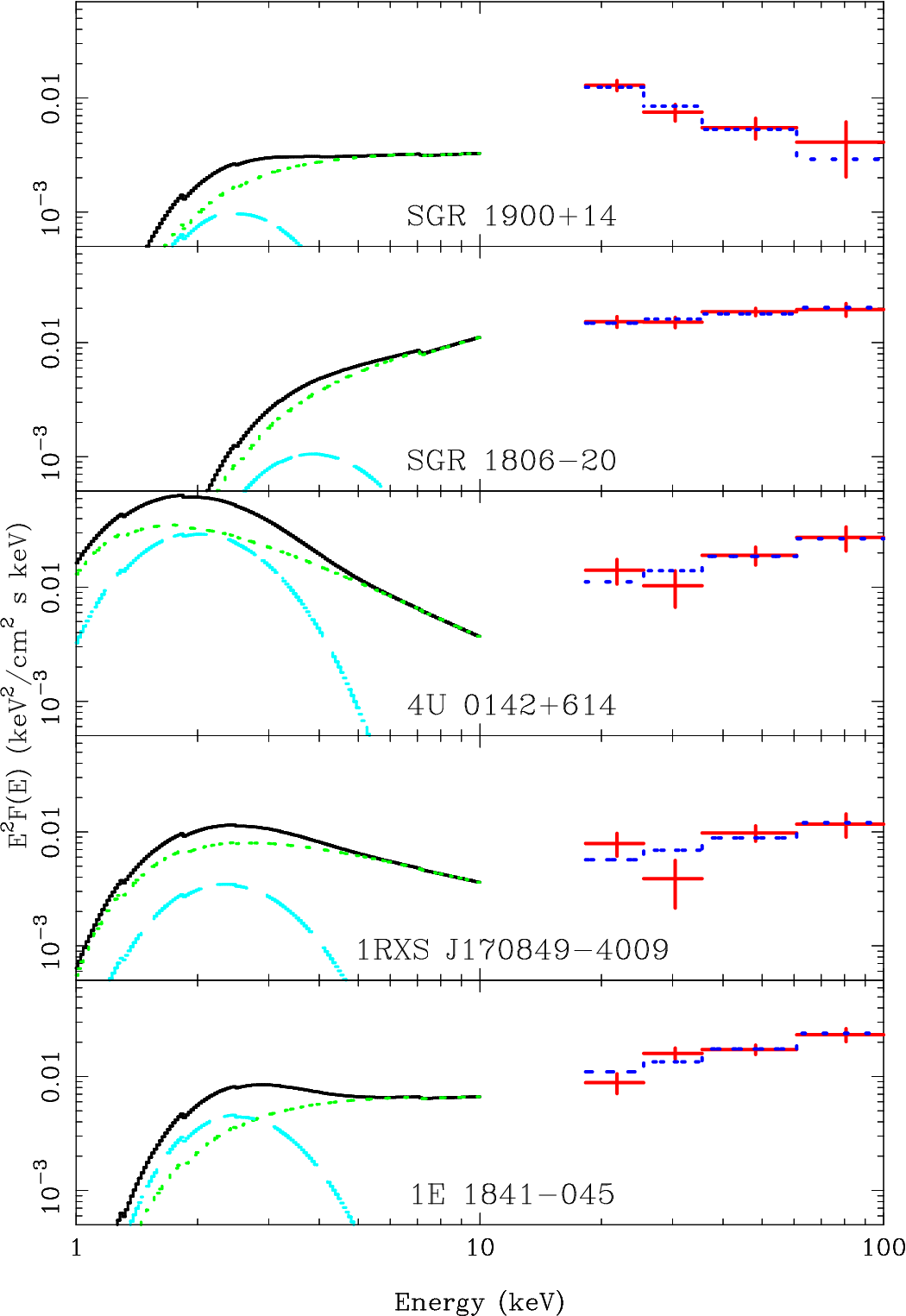} 
\caption{Broad band X--ray spectra of the five magnetars detected by \int~. The
data points above 18 keV are the \int~ spectra with their best fit
power-law models (dotted lines). The solid lines below 10 keV represent
the absorbed power-law (dotted lines) plus blackbody (dashed lines) models
taken from \cite{woods01} (\zerozero , during a quiescent state in spring 2000), \cite{xmm1806} (\zerosei, observation B, when the bursting activity was low), \cite{gohler} (4U~0142+614), \cite{rea}
(1RXS~J170849--4009), and \cite{morii} (1E~1841--045). From \cite{gotz1900}.}
\label{fig:bbsp}
\end{figure}

In order to coherently compare the broad band spectral properties
of all the SGRs and AXPs detected at high energy, we analysed all
the public \int~ data using the same procedures. Our results are shown
in Fig. \ref{fig:bbsp}, where the \int~ spectra are plotted together
with the results of observations at lower energy taken from the
literature (see figure caption for details).

As can be seen, AXPs generally present harder spectra than SGRs in hard X-rays. In particular, for the three
AXPs, a spectral break is expected to occur between 10 and 20 keV in order to reconcile the soft and the 
hard parts of the spectrum. On the other hand, SGRs present a softer spectrum at higher energies, also implying
a break around 15 keV (especially for \zerozero), but in the opposite sense with respect to the AXPs. The fact that
the spectral break is more evident in \zerozero~ could be due to the fact that its level of activity was much lower
during our observations, compared to \zerosei. All three AXPs, on the other hand, can be considered being in a quiescent state since no bursts were been reported from them during the \int~ observation.

The magnetar model, in its different flavours, explains this hard X-ray emission as powered by bremsstrahlung photons
produced either close to the neutron star surface or at a high altitude ($\sim$ 100 km) in the magnetosphere
\cite{tlk,tb05}.  The two models could be distinguished by the presence of a cutoff at $\sim$ 100 keV or $\sim$ 1 MeV. 
Unfortunately current experiments like \int~  are not sensitive enough to firmly assess the
presence of the cutoffs and hence to distinguish between the two models.

\section{Conclusions}

Thanks to \int, and in particular to its imager IBIS, we have been able to study most of the magnetars' phenomenology
with unprecedented sensitivity at high energies. One of the most striking results is the discovery, which
was particularly unexpected for AXPs, of the persistent hard X-ray emission. This discovery, which
can be considered one of the most important \int~ results, represents a new important input for theoreticians
who started to include it in the magnetar model (see e.g. \cite{belo}). 

Also, the fact that short bursts evolve with time is a new feature that has to be considered with care within the 
magnetar model: up to now no clear explanation has been provided for this.

The large number of detected short bursts from \zerosei~ allowed us a good
determination of the shape and slope of their Number-Intensity distribution, showing that a single
power law holds over 2.5 orders of magnitude. The faint end of the distribution represents the faintest bursts ever detected at these energies. 

In addition, the fact that \zerosei~ has been particularly active during the last years, also emitting a 
once-in-a-lifetime event such as the giant flare (and its possible high energy afterglow), has allowed  
observations of relatively rapid changes of the bursting and persistent emission of a Magnetar and   
to interpret them with the evolution of a very strong and complicated magnetic field, confirming
the magnetic field as the dominant source of energy in Soft Gamma-Ray Repeaters and Anomalous X-ray
Pulsars. Our group will continue the monitoring program of SGRs in order to better understand these peculiar objects.

\section*{Acknowledgements}
DG acknowledges the French Space Agency (CNES) for financial support. Based on observations with INTEGRAL, an ESA project with instruments and the science data centre funded by ESA member states (especially the PI countries: Denmark, France, Germany, Italy, Switzerland, Spain), Czech Republic and Poland, and with the participation of Russia and the USA. ISGRI has been realized and maintained in flight by CEA-Saclay/DAPNIA with the support of CNES.


\begin{thebibliography}{}
\bibitem{belo} Belobodorov, A.~M., 2006, Proc. ``Isolated Neutron Stars: from the Interior to the Surface", astro-ph/0608372
\bibitem{flare} Borkowski, J., G\"{o}tz, D., Mereghetti, S., et al., 2006, GCN, 2920
\bibitem{dt92} Duncan, R.C., \& Thompson, C., 1992, ApJ 392, L9
\bibitem[Esposito et al. (2006)]{esposito06}Esposito, P., Mereghetti, S., Tiengo, A., et al., 2006, A\&A, in press, astro-ph/0609078
\bibitem[Gavriil (2002)]{gavrill}Gavriil, F., Kaspi, V.M., \& Woods, P.M., 2002, Nature, 419, 142
\bibitem[G{\"o}hler et al. (2005)]{gohler} G{\"o}hler, E., Wilms, J., \& Staubert, R., 2003, A\&A 433, 1079
\bibitem{gotz04}G\"{o}tz, D., Mereghetti, Mirabel, F.I., \& Hurley, K., 2004, A\&A 417, L45
\bibitem[G\"otz et al. (2006)]{gotz1806}G\"{o}tz, D., Mereghetti, S., Molkov, S., A\&A, 2006a, 445, 313 
\bibitem[G\"{o}tz et al. (2006)]{gotz1900}G\"{o}tz, D., Mereghetti, S., Tiengo, A., et al., 2006b, A\&A 449, L31
\bibitem{granot}Granot J., Ramirez-Ruiz, E., Taylor, G.~B., et al., 2006, ApJ 638, 391
\bibitem{denhartog}den Hartog, P.~R., Hermsen, W., Kuiper, L., et al., 2006, A\&A, in press, astro-ph/0601644
\bibitem[Hurley et al. (1999)]{hurley1999}Hurley, K., Cline, T., Mazets, E., et al., 1999, Nature 397, 41
\bibitem{kh1900}Hurley, K., Li, P., Kouveliotou, C., et al., 1999, ApJ, 510, L111
\bibitem{ipn1900}Hurley, K., Mazets, E., Golenetskii, S., et al., 2002 GCN Circ. 1715
\bibitem{hurley2003} Hurley, K., Atteia, J.-L.,  Kawai, N., et al., 2003, GCN, 2308
\bibitem[Hurley et al. (2005)]{rhessigiant} Hurley, K., Boggs, S. E., Smith, D. M., et al., 2005, Nature 434, 1098 
\bibitem{ibrahim}Ibrahim, A.~I., Strohmayer, T. E., Woods, P. M., et al., 2001, ApJ 558, 237
\bibitem{kuiper}Kuiper, L.,Hermsen, W., den Hartog, P.~R. et al., 2006, ApJ, 645, 556
\bibitem{giallo} Israel, G., Covino, S., Mignani, R., et al., 2005, A\&A 438, L1
\bibitem[Kaspi (2003)]{kaspi} Kaspi, V.M., Gavriil, F.P., Woods, P.M., et al., 2003, ApJ, 588, L93
\bibitem[Kaspi (2006)]{kaspi4u} Kaspi, V., Dib, R., \& Gavriil, F., 2006, ATel 794 
\bibitem[Kouveliotou et al. (1998)]{kouve98}Kouveliotou, C., Dieters, S., Stromayer, T., et al., 1998, Nature 393, 235
\bibitem{1806disc} Laros, J., Fenimore, E. E., Fikani, M. M., et al., 1986 Nature 322, 152
\bibitem{lenters}Lenters, G.~T., Woods, P. M., Goupell, J. E., 2003, ApJ 587, 761 
\bibitem[Mattana et al. (2006)]{mattana} Mattana, F., G\"{o}tz, D., Falanga, M., et al. 2006, A\&A, in press
\bibitem[Mazets et al. (1979)]{mazets}Mazets, E.~P., Golentskii, S. V., Ilinskii, V. N., et al., 1979a Nature 282, 587
\bibitem{1900discovery}Mazets, E.~P., Golenetskij, S. V., \& Guryan, Y. A., et al., 1979b Sov. Astr. Lett 56, 343 
\bibitem{mazets05} Mazets E.P., Cline, T.L., Aptekar, R.L., et al. 2005, astro-ph/0502541
\bibitem[Mereghetti \& Stella (1995)]{merestella}Mereghetti, S. \& Stella, L., 1995, ApJ, 442 L17
\bibitem[Mereghetti (2002)]{mere02}Mereghetti, S., Chiarlone, L., Israel, G.L., Stella, L., 2002, Proc 270 WE-Heraeus Seminar, 29
\bibitem[Mereghetti et al. (2005a)]{acsgiant}Mereghetti S., G\"{o}tz, D., von Kienlin, A., et al. 2005a, ApJ, 624, L105
\bibitem[Mereghetti et al. (2005b)]{xmm1806}Mereghetti S., Tiengo, A., Esposito, P., et al., 2005b, ApJ 628, 938
\bibitem{mere05} Mereghetti S. , G\"{o}tz, D., Mirabel, I.F., \& Hurley, K., 2005c, A\&A 433, L9
\bibitem[Mereghetti et al. (2006a)]{xmm1900}Mereghetti S., Tiengo, A., Esposito, P., et al., 2006a, ApJ, in press,
\bibitem[Mereghetti et al. (2006b)]{mere2006}Mereghetti S., Esposito, P., Tiengo, A., 2006b, Proc. ``Isolated Neutron Stars: from the Interior to the Surface", astro-ph/0608364
\bibitem{molkovaxp}Molkov, S.~V., Cherepashchuk, A.~M., Lutovinov, A.~A., et al., 2004, Astronomy Letters 30, 534
\bibitem{molkov}Molkov, S., Hurley, K., Sunyaev, R., et al., 2005, A\&A 433, L13 
\bibitem[Morii et al. (2003)]{morii} Morii, M., Sato, R., Kataoka, J., \& Kawai, N., 2003, PASJ 55, L45 
\bibitem{pac92} Paczynski, B., 1992, AcA 42, 145 
\bibitem[Palmer et al. (2005)]{swiftgiant}Palmer, D.~M., Barthelmy, S., Gehrels, N., et al., 2005, Nature 434, 1107
\bibitem{swift1900}Palmer, D.~M., Sakamoto, T., Barthelmy, S., et al., 2006a,  ATel 789
\bibitem{1900outburst}Palmer, D.~M., Sakamoto, T., Barthelmy, S., et al., 2006b, GCN Circ. 4949
\bibitem[Rea et al. (2005)]{rea} Rea, N., Oosterbroek, T., Zane, S., et al., 2005, MNRAS 361, 710
\bibitem{revnitsev}Revnivtsev, M.~G., Sunyaev, R.~A., Varshalovich, D.~A., et al., 2004, Astronomy Letters 30, 382 
\bibitem{tera}Terasawa, T., Tanaka, Y.~T., Takei, Y., et al., 2005, Nature 434, 1110
\bibitem{td95}Thompson, C., \& Duncan, R.C., 1995, MNRAS 275, 255
\bibitem[Thompson et al. (2002)]{tlk}Thompson, C., Lyutikov, \& M. Kulkarni, S. R., 2002, ApJ, 574, 332
\bibitem{tb05}Thompson, C., \& Beloborodov, A.~M. 2005, ApJ 634, 565
\bibitem[Wang et al. (2006)]{wang} Wang, Z., Chakrabarty, D., \& Kaplan, D.~L., 2006, Nature, 440, 772
\bibitem{woods99} Woods, P.~M., Kouveliotou, C., van Paradijs, J., et al., 1999, ApJ 519, L139, 
\bibitem[Woods et al. (2001)]{woods01}Woods, P.~M., Kouveliotou, C., G{\"o}{\u g}{\"u}{\c s}, E., et al., 2001,  ApJ 552, 748
\bibitem[Woods et al. (2005)]{woods} Woods, P.M., Kouveliotou, C., Gavriil, F.P., et al. 2005, ApJ, 629, 985



\end{thebibliography}

\end{document}